\documentclass[12pt]{article}
\usepackage{amssymb,amsmath,latexsym}
\usepackage{graphicx}
\begin{document}
\title{CHAOS IN FRACTIONALLY INTEGRATED GENERALIZED AUTOREGRESSIVE CONDITIONAL HETEROSKEDASTIC PROCESSES}
\date{\vspace{-5ex}}

\author{A.YILMAZ and G. UNAL}
\maketitle
\begin{center}

Financial Economics Graduate Program, Yeditepe University
34755, Istanbul, Turkey
adil.yilmaz@std.yeditepe.edu.tr, gunal@yeditepe.edu.tr

\end{center}

\abstract{Fractionally integrated generalized autoregressive conditional heteroskedasticity (FIGARCH) arises in modeling of financial time series. FIGARCH is essentially governed by a system of nonlinear stochastic difference equations.
\[
{u_t}={z_t}
\]
\[
(1-\sum\limits_{j=1}^q \beta_j L^j)\sigma_{t}^2=\omega+(1-\sum\limits_{j=1}^q \beta_j L^j-(\sum\limits_{k=1}^p \varphi_k L^k)(1-L)^d)u_{t}^2
\]
where $\omega\in$ R, and $\beta_j\in$ R are constant parameters, $\{u_t\}_{{t\in}^+}$ and $\{\sigma_t\}_{{t\in}^+}$ are the discrete time real valued stochastic processes which represent FIGARCH (p,d,q) and stochastic volatility, respectively. Moreover, L is the backward shift operator, i.e. $L^d u_t \equiv u_{t-d}$ (d is the fractional differencing parameter 0$<$d$<$1).\\
\indent
In this work, we have studied the chaoticity properties of FIGARCH (p,d,q) processes by computing mutual information, correlation dimensions, FNNs (False Nearest Neighbour), the Lyapunov exponents, and for both the stochastic difference equation given above and for the financial time series. We have observed that maximal Lyapunov exponents are negative, therefore, it can be suggested that FIGARCH (p,d,q) is not deterministic chaotic process.
}\\
\\

\section{Introduction}

Detection of chaotic behavior in financial and economic (both micro and macro) data has been the topic of numerous scientific studies such as (Dechert and Gençay, 2000; Das and Das, 2006-7; Moeni et al., 2007; Günay, 2015). Existence of chaos in data favors the short-term predictability and controllability (Abarbanel, 1996) of the underlying difference equations which draws the attention of the scientific circles. However, when dealing with financial and economic data one should always bear in mind that GARCH (Generalized autoregressive conditional heteroscedasticity) models (Francq and Zaqoian, 2010) mimic the stylized facts. The latter is a set of nonlinear stochastic difference equations, therefore, it is quite challenging idea to associate it with deterministic chaos. Here we focus on the chaoticity properties of FIGARCH (Fractionally Integrated generalized autoregressive conditional heteroscedasticity) model by considering correlation dimension and Lyapunov exponents.\\
\indent
FIGARCH model was introduced by Baillie, Bollerslev and Mikkelsen (1996) by modifying GARCH model to provide more persistence on the conditional variance. The model allows slow hyperbolic rate of decay for the innovations in the conditional variance and it has an ability to estimate long memory of conditional volatility. Recent researches have found evidence of long-range dependence for a variety of financial assets and strong evidence of long memory in volatility (Cujaeiro, D. O. et al. 2008).\\
\indent
Since its discovery four decades ago, chaos theory has been attracting a lot of interest.  The existence of chaotic behavior has been studied in many types of disciplines, ranging from atmospheric dynamics (e.g. Lorenz, 1969; Essex et al., 1987), geophysics (e.g. Hense, 1987; Wilcox et al., 1991; Lorenz, 1996; Sivakumar, 2004),  medicine (e.g. Almog et al., 1990; Goldberger et al. 1988; Babloyantz, 1985; Sviridova et al., 2015), turbulence (e.g. Abarbanel, 1994), to financial markets (e.g. Hsieh, 1991; DeCoster, 1992; Cornelis, 2000; Frezza, 2014), and electrical circuits (e.g. Yim et al., 2004).\\
\indent
Chaotic systems are deterministic systems which are unpredictable in the long term due to their sensitivity to even a very small change of initial conditions. In the deterministic picture, irregularity can be autonomously generated by the nonlinearity of the intrinsic dynamics. The most direct link between chaos theory and the real world is the analysis of time series data in terms of nonlinear dynamics. Chaos theory has inspired a new set of useful time series tools and provides a new language to formulate time series problems (Schreiber T., 1998).\\
\indent
This paper investigates the existence of chaoticity in nonlinear FIGARCH model by using simulated time series and nonlinear difference equation directly. As a start point, assuming that FIGARCH is a deterministic chaotic system, as a common practice, phase space is reconstructed by employing delay coordinate embedding technique by Takens’ theorem which justifies that with appropriate embedding dimension and delay time, the reconstructed phase space is the one to one image of the original system and has got the same mathematical properties.\\
\indent
Therefore, in third section embedding dimension and delay time are determined. In order to estimate appropriate time delay, the mutual information method is applied. Then, embedding dimension is found out by false nearest neighbor method. The correlation dimension provides a tool to quantify self-similarity. Therefore, in the next section, correlation dimension is calculated by employing Grassberger and Procaccia’s procedure. In fifth section, Lyapunov exponent is calculated to quantify the sensitivity to initial conditions which is the most essential characteristic of chaos. For this purpose, different algorithms are employed. Wolfs' and Kantz’s algorithms are employed to simulated time series as well as utilizing more direct method by constructing dimensional map from difference equation. The summary and conclusions of this paper are presented in Section 6. 
\section{FIGARCH (Fractionally Integrated Generalized Autoregressive Conditional Heteroskedasticity)}

Generalized Autoregressive Conditional Heteroscedastic (GARCH) model and Integrated GARCH (IGARCH) model were developed by Bollerslev (1986) and Engle and Bollerslev (1986) respectively. GARCH model suffers from several problems, such as non-negativity problem and issue with leverage effects. Besides, the model doesn’t allow for any direct feedback between the conditional variance and the conditional mean. On the other side, in most of the empirical situations, the IGARCH model seems to be too restrictive as it implies infinite persistence of a volatility shock. \\
\indent
Inspired by these problems, Fractionally Integrated GARCH (FIGARCH) model was introduced by Baillie,Bollerslev, and Mikkelsen (1996) as a new process, generalizing the well-known GARCH to allow persistence in the conditional variance. It was developed for the purpose of a more flexible class of processes for the conditional variance that are more capable of explaining and representing the observed temporal dependencies in financial market volatility (Baillie, Bollerslev and Mikkelsen,1996).\\
\indent
FIGARCH model is simply obtained by replacing the first difference operator with fractional differencing operator in GARCH (p, q). So, FIGARCH(p, d, q) is written as,
\begin{equation}
\phi(L)(1-L)^d u_{t}^2=\omega+[1-\beta(L)]\varepsilon_{t}
\end{equation}
where 0$<$d$<$1,and all the roots of $\phi(L)$ and $[1-\beta(L)]$ lie outside the unit circle. If FIGARCH(p,d,q) can be rearranged as;
\begin{equation}
[1-\beta(L)]\sigma_{t}^2=\omega+[1-\beta(L)]-\phi(L)(1-L)^d]u_{t}^2
\end{equation}
\begin{equation}
(1-\sum\limits_{j=1}^q \beta_j L^j)\sigma_{t}^2=\omega+(1-\sum\limits_{j=1}^q \beta_j L^j-(\sum\limits_{k=1}^p\phi_k L^k)(1-L)^d)u_{t}^2
\end{equation}
So, the conditional variance of $u_t$ is given by,
\begin{equation}
\sigma_{t}^2=\omega[1-\beta(L)]^{-1}+\{1-[1-\beta(L)]^{-1}\phi(L)(1-L)^d\}u_{t}^2
\end{equation}
\begin{equation}
\sigma_{t}^2\equiv\omega[1-\beta(L)]^{-1}+\lambda(L)u_{t}^2
\end{equation}

where $\lambda(L)=\lambda_1 L+\lambda_2 L^2+...$.Of course, for the FIGARCH(p, d, q), for (8) to be well-defined, the conditional variance in the ARCH($\infty$) representation in (10) must be non-negative, i.e., $\lambda_k\geq0$ for k = 1, 2, ...  (Baillie et al., 1996).\\
\indent
Conrad and Haag (2006) also introduced a set of conditions that guarantees the non-negativity of the conditional variance in all situations. Moreover, Davidson (2004) had shown that FIGARCH model possesses more memory than a GARCH or IGARCH model.
\section{CHAOTIC BEHAVIOR and FIGARCH}
Chaotic behavior is irregularity of motion, unpredictability and sensitivity to initial conditions. Our purpose is to identify FIGARCH nonlinear stochastic difference equation’s chaotic properties in this sense. Its chaotic nature will basically give us an idea about its predictability horizon and forecasting quality.\\
\indent
In order to measure chaos of time series data, Lyapunov exponent can be estimated which is a measure of the average speed with which infinitesimally close states separate. To be able to find out Lyapunov exponents of the data, first it is needed to go from scalar time series data to the multivariate state or phase space which is required for chaotic motions to occur in the first place.

\subsection{Reconstructing Phase Space}

The answer of the question of how to go from scalar to multivariate state is the geometric theorem called the embedding theorem attributed to Takens and Mane (1981). Abarnel asserts that all variables in a nonlinear process is generically connected and they influence each other. \\
\indent
Considering that we have time series data of $\{x_0, x_1,...x_i,...x_n\}$, Takens implies that the reconstructed attractor of the original system can be written as the vector sequence
\begin{equation}
p(i)=(x_i, x_{i+T}, x_{i+2T},...x_{i+(d-1)T})
\end{equation}

where T represents the embedding delay and d represents embedding dimension.\\
\indent
Takens also states that for a large enough d, many important properties of the original system are reproduced without ambiguity in the new space of vectors. In other words, the attractor constructed will have the same mathematical properties as the original system (such as dimension, Lyapunov exponents etc.).\\
\indent
As it is seen from the formula, in order to be able to reconstruct the attractor, proper values of the embedding delay and embedding dimension must be determined.\\

\textbf{Data}

In order to analyze FIGARCH data, several FIGARCH(1,d,1) models are generated. While alpha, beta and omega coefficients remained unchanged at 0.01 to be able to keep ranging d values from 0.05 to 0.9 with 8192 number of data points. For each d value, 50 random samples are generated and results are tested. Each model are simulated by using Kevin Sheppards’s MFE Toolbox and then for verification, as a second source Oxmetrics Garch package is applied.

\subsection{Mutual Information}

Embedding delay (T) value is determined by looking for the first minimum of the nonlinear correlation function named mutual information. The mutual information is introduced by Fraser and Swinney (1986), between $x_i$ and $x_{i+T}$ as a suitable quantity for determining T.\\
\indent
There are two important principles in estimation of T. 
\begin{enumerate}
  \item 	T has to be large enough so that the information in $x_{i+T}$ is significantly different from the information in $x_i$.
  \item 	T shouldn’t be too large that $x_{i+T}$ and $x_i$ are completely independent in statistical sense.
\end{enumerate}
Giving sets $A=\{a_i\}$ and $B=\{b_j\}$ mutual information between them in bits written as; 
\begin{equation}
log_2 \Bigg[\frac{P_{AB} (a_i, b_j)}{P_A(a_i) P_B(b_j)}\Bigg]
\end{equation}
where $P_A (a_i)$ and $P_B (b_j)$ are the individual probability densities, while $P_{AB} (a_i,b_j)$ is the joint probability density of A and B.\\
\indent
If $a_i$ and $b_j$ is completely independent $P_{AB} (a_i,b_j )=P_A (a_i) P_B (b_j)$ and the mutual information is zero. The average of all measurements of information statistic between A and B measurements is written as;
\begin{equation}
I_{AB}=\sum\limits_{a_i, b_j}P_{AB} (a_i,b_j )log_2 \Bigg[\frac{P_{AB} (a_i, b_j)}{P_A(a_i) P_B(b_j)}\Bigg]
\end{equation}
Mutual information measures mutual dependence of two sets based on the notion of the information between them.
So, for the measurements s(t) at time t which are connected to the measurement s(t+T), if we rewrite; 
\begin{equation}
I(T)=\sum\limits_{s(n), s(n+T)} P(s(n),s(n+T))log_2 \Bigg[\frac{P(s(n),s(n+T))}{P_A(s(n))P_B(s(n+T))}\Bigg]
\end{equation}
when $T\to\infty$, $I(T)\to0$ since correlation between s(n) and s(n+T) disappears (Abarbanel, 1996).\\
\indent
Fraser suggests that as I(T) is a kind of autocorrelation function, therefore it is appropriate to choose time delay value at first minimum of the mutual information although it could sometimes be misleading considering linear nature of the method.\\
\indent
For each model, in order to identify proper embedding delay, mutual information of models are computed and results are found as in Table 1.
\begin{figure}[!ht]
  \centering
    \includegraphics[width=0.85\linewidth]{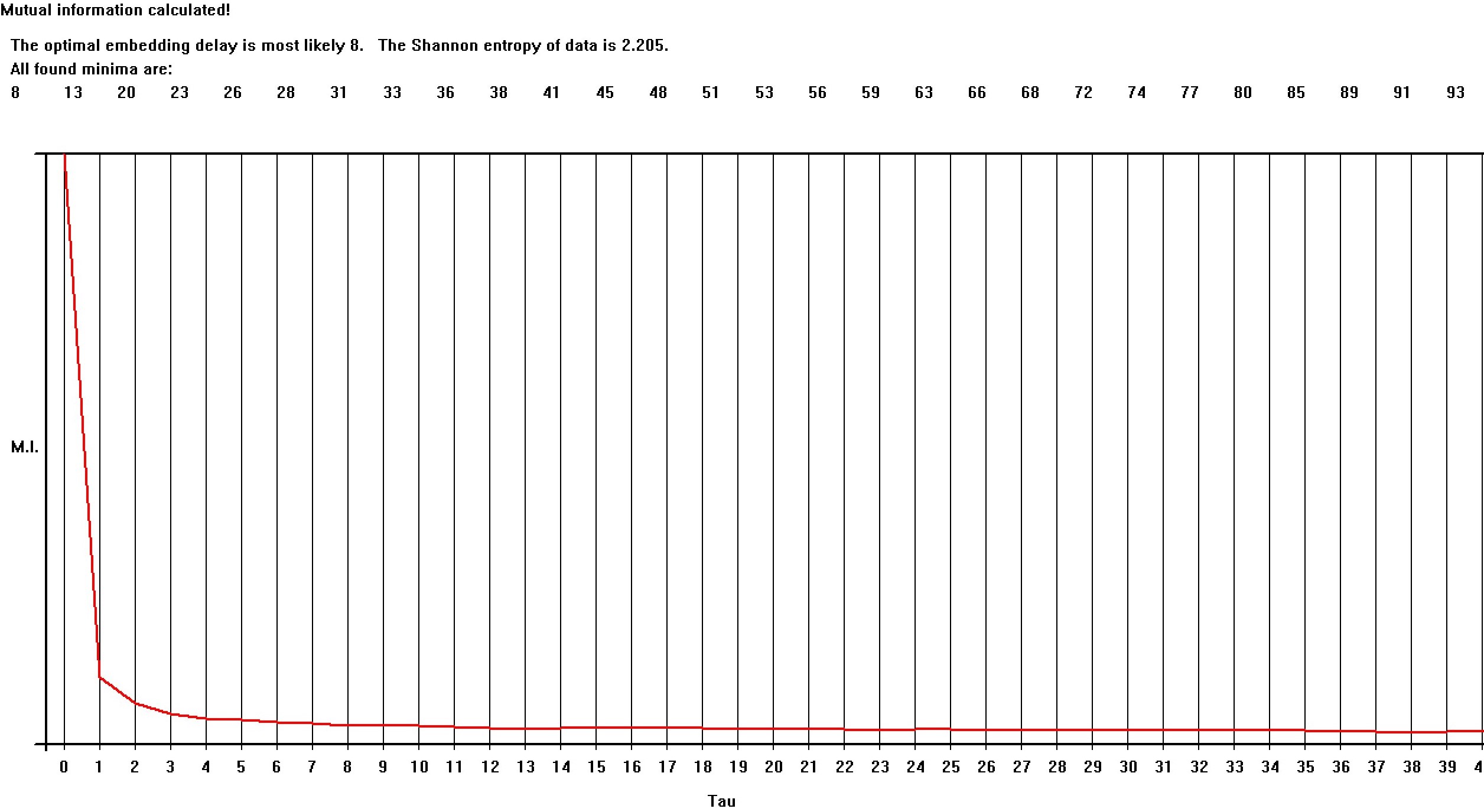}
      \caption{Mutual Information for Figarch d=0.90.}
\end{figure}
\begin{table}[h!]
\centering
\begin{tabular}{|c|c|} 
 \hline
 Figarch Model & Embedding delay \\ [0.5ex] 
 \hline
 Figarch d=0.05 & 2 \\ 
 Figarch d=0.15 & 1 \\
 Figarch d=0.25 & 2 \\
 Figarch d=0.35 & 4 \\
 Figarch d=0.45 & 6 \\ 
 Figarch d=0.55 & 7 \\ 
 Figarch d=0.65 & 8 \\
 Figarch d=0.75 & 8 \\ 
 Figarch d=0.80 & 6 \\
 Figarch d=0.90 & 8 \\ [0.5ex]
 \hline
\end{tabular}
\caption{Mutual Information for each Figarch Model.}
\end{table}
 \subsection{False Nearest Neighbor Method}
Supposing that a state space reconstruction is made in dimension d with data vectors using the time delay suggested by mutual information.
\begin{equation}
y(i)=(x_i,x_{i+T},x_{i+2T},...,x_{i+(d-1)T} )
\end{equation}
The nearest neighbor in phase space will be a vector;

\begin{equation}
y^{NN}(i)=(x_i^{NN},x_{i+T}^{NN},x_{i+2T}^{NN},...,x_{i+(d-1)T}^{NN} )
\end{equation}

If the vector $y^{NN} (i)$ is a false neighbor of $y(i)$ having arrived its neighborhood by projection from a higher dimension because the present dimension d doesn't unfold the attractor, then by going to next dimension $d+1$ this false neighbor may be removed out of the neighborhood of $y(i)$.\\
\indent
By looking at every data point $y(i)$  and asking at what dimension all false neighbors are removed, we will sequentially intersections of orbits of lower and lower dimension are removed until at last point intersections are removed. At that juncture d will have been identified where the attractor is unfolded.\\
\indent
Comparing the distance between the vectors $y(i)$ and $y^{NN} (i)$ in dimension $d$ with the distance between the same vectors in dimension $d + 1$, it can easily be established which are true neighbors and which false. It only needs to be compared  $x_{(i+dT)}-x_{(i+dT)}^{NN}$  with the Euclidian distance $| y_i-y_i^{NN} |$ between nearest neighbors in dimension $d$. \\
\indent
If the additional distance is large compared to the distance in dimension $d$ between nearest neighbors, then we have a false neighbor.\\
\indent
The square of the Euclidian distance between the nearest neighbor points as seen in dimension d is
\begin{equation}
R_d (i)^2=\sum\limits_{m=1}^d [x_{i+(m-1)T}-x_{i+(m-1)T}^{NN}]^2
\end{equation}
while dimension d+1 it is; 
\begin{equation}
R_{d+1} (i)^2=\sum\limits_{m=1}^{d+1} [x_{i+(m-1)T}-x_{i+(m-1)T}^{NN}]^2
\end{equation}
\begin{equation}
R_{d+1} (i)^2=R_d (i)^2 + |x_{i+dT}-x_{i+dT}^{NN}|^2
\end{equation}
The distance between points when seen in dimension $d+1$ relative to the distance in dimension $d$ is;
\begin{equation}
\sqrt{\frac{R_{d+1}(i)^2 -R_d(i)^2}{R_d(i)^2}}=\frac{x_{i+dT} -x_{i+dt}^{NN}}{R_d(i)}>r_{tol}
\end{equation}
When this quantity is larger than some threshold, we have a false neighbor (Kennel M B, Brown R and Abarbanel H D 1992).\\
\indent
Plot of percentage of false neighbors show the unfolded geometry and where there is no unfolding any more. With the correct choice of $d$ dimension, modelling the data in $d$ number of dynamical degrees of freedom will be adequate to capture the properties of the source.\\ 
\begin{figure}[!ht]
  \centering
    \includegraphics[width=0.85\linewidth]{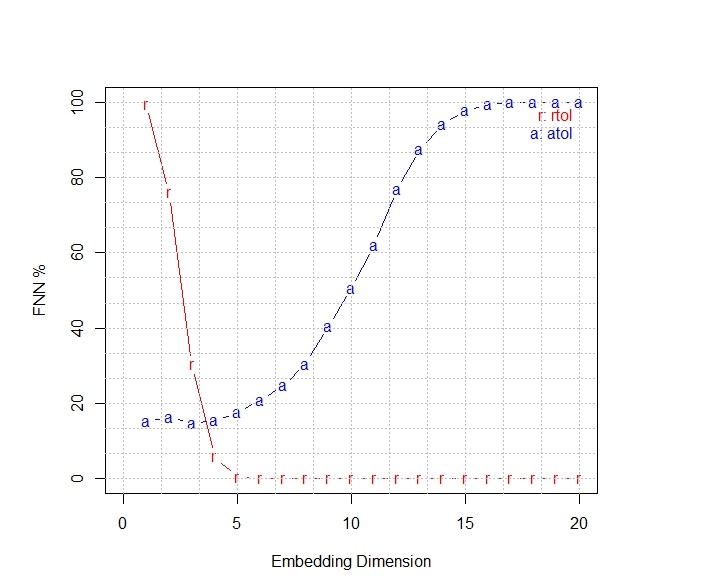}
      \caption{FNN embedding dimension result for FIGARCH d=0.90.}
\end{figure}
\indent
The figure 2 shows minimum embedding dimension where percentage of nearest neighbors goes to zero taken into account some threshold $r_{tol}$. Disappearance of false neighbors indicates minimum embedding dimension $r_{tol}$ is false neighbor Euclidian distance tolerance and $a_{tol}$ is neighbor tolerance based on attractor size. The neighbors are declared false neighbors, when the ratio of the Euclidian distances between neighbor candidates in successive embedding dimensions exceeds $r_{tol}$.
\begin{table}[h!]
\centering
\begin{tabular}{|c|c|c|} 
 \hline
 Figarch Model & Embedding Delay & Embedding Dimension \\ [0.5ex] 
 \hline
 Figarch d=0.05 & 2 & 6 \\ 
 Figarch d=0.15 & 1 & 7 \\
 Figarch d=0.25 & 2 & 6 \\
 Figarch d=0.35 & 4 & 8 \\
 Figarch d=0.45 & 6 & 6\\ 
 Figarch d=0.55 & 7 & 6\\ 
 Figarch d=0.65 & 8 & 6\\
 Figarch d=0.75 & 8 & 6\\ 
 Figarch d=0.80 & 6 & 6\\
 Figarch d=0.90 & 8 & 6\\ [0.5ex]
 \hline
\end{tabular}
\caption{FNN embedding dimension results for each FIGARCH model}
\end{table}
\section{Correlation Dimension}
Correlation dimension is a widely used and accepted tool to analyze degree of complexity. Grassberger and Procaccia (1983) was introduced a useful method in order to compute correlation dimension which measure an attractor dependent on a contraction rate of a fractal measure in some phase space in a given set. They defined correlation sum which approximates the probability of having pair of points with separation distance less than a given size $\varepsilon$ as,
\begin{equation}
C(\varepsilon)=\frac{1}{N^2}\sum\limits_{i,j}^N \Theta (\varepsilon-\|x_i - x_j\|)
\end{equation}
where $\Theta$ is the Heaviside step function,
\begin{equation}
\begin{cases} 
      \Theta(\varepsilon-\|x_i - x_j\|)=1 & 0\leq(\varepsilon-\|x_i - x_j\|) \\
      0 & (\varepsilon-\|x_i - x_j\|)<0 \\
   \end{cases}
\end{equation}
When $N\to\infty$, for small values of $\varepsilon$, C follows a power law;
\begin{equation}
C(\varepsilon)\propto \varepsilon^{D_C}
\end{equation}
where $D_C$ is the correlation dimension. Therefore, $D_C$ is defined as;
\begin{equation}
D_C=\lim_{\varepsilon\to0} \lim_{N\to\infty} \frac{\partial lnC(\varepsilon)}{\partial ln(\varepsilon)}
\end{equation}
\indent
Because of the slow convergence, slope of the straight line by using least square fit in a plot of $lnC(\varepsilon)$  vs. $ln(\varepsilon)$ was computed for the estimation of correlation dimension in this work. \\
\indent
Correlation dimension is calculated for each Figarch $d$ value simulation for embedding delays (from 1 to 20) and embedding dimensions from 1 to 20 (Figure 3).\\ 
\begin{figure}[!ht]
  \centering
    \includegraphics[width=0.85\linewidth]{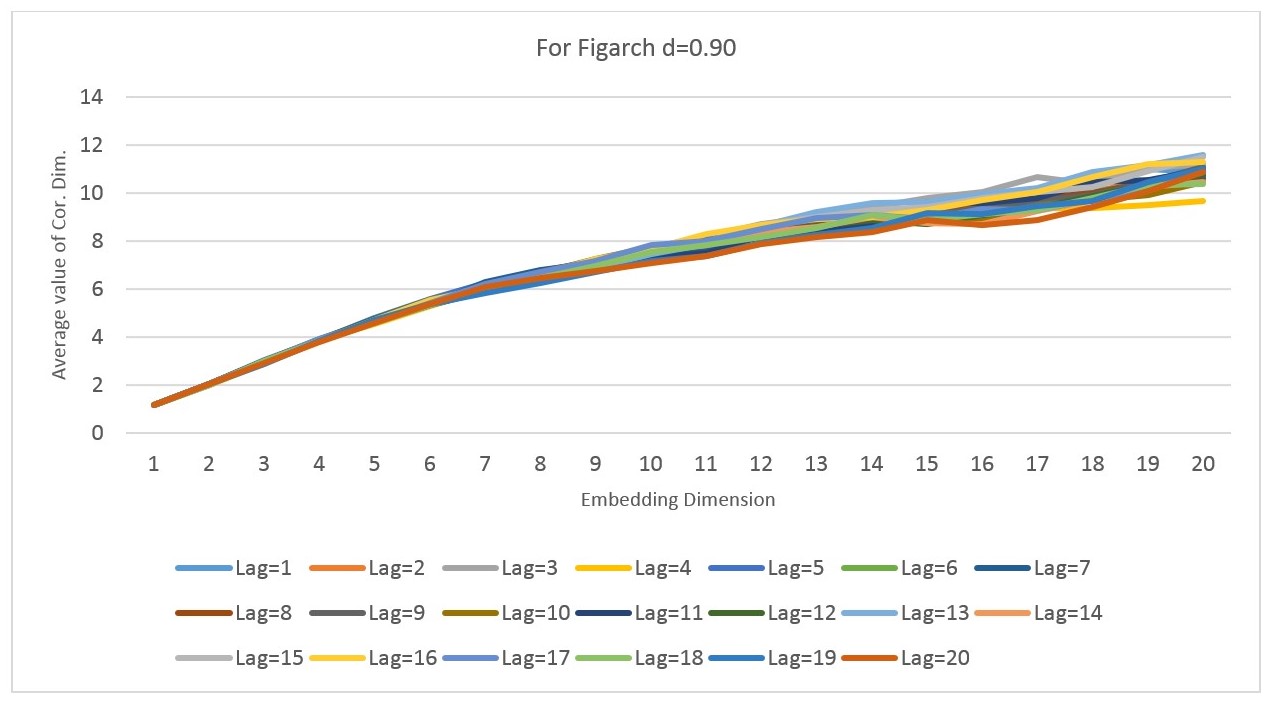}
      \caption{Correlation dimension results for embedding delays from 1 to 20 and embedding dimensions from 1 to 20 for each FIGARCH model.}
\end{figure}
\indent
According to determined embedding delay values by mutual information estimation, correlation dimension-embedding dimension results appeared in Figure 4. Correlation dimension value doesn't converge in any of them. So, embedding dimension can't be determined.\\
\begin{figure}[!ht]
  \centering
    \includegraphics[width=0.85\linewidth]{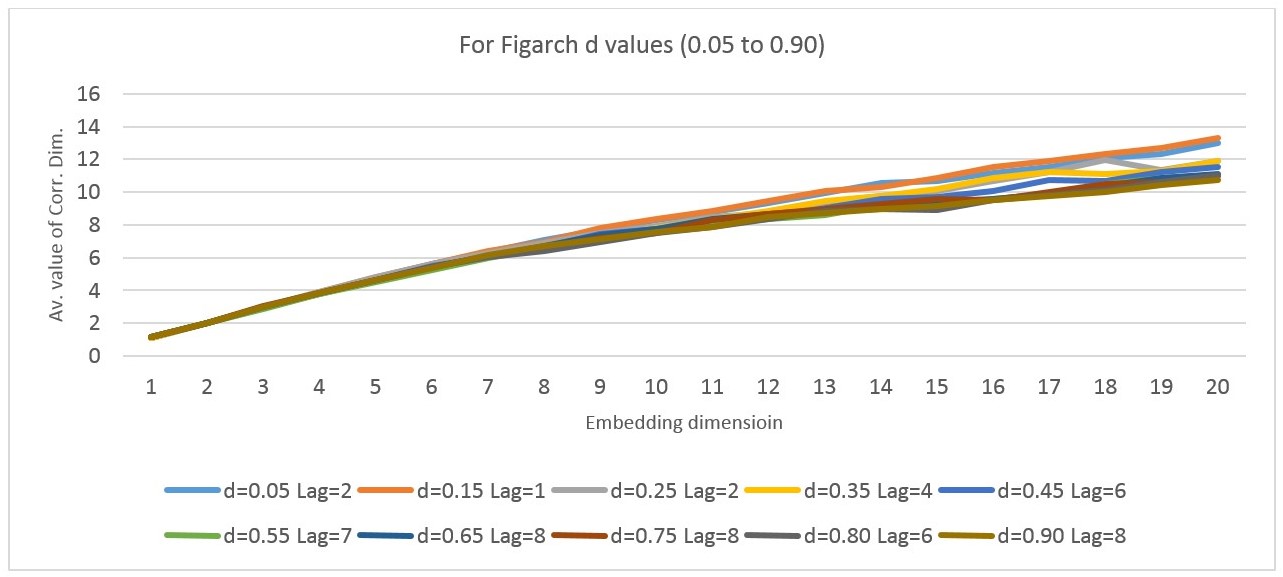}
      \caption{Correlation dimension results according to embedding delay values determined by mutual information for each FIGARCH model.}
\end{figure}
\indent
Then, by assuming, for different embedding delay values, there might be a convergence in correlation dimension values, for several embedding delay values (1 to 20) and for embedding dimensions from 1 to 20, correlation dimension values are calculated. Again, no clear convergence is observed.\\   
\begin{figure}[!ht]
  \centering
    \includegraphics[width=0.85\linewidth]{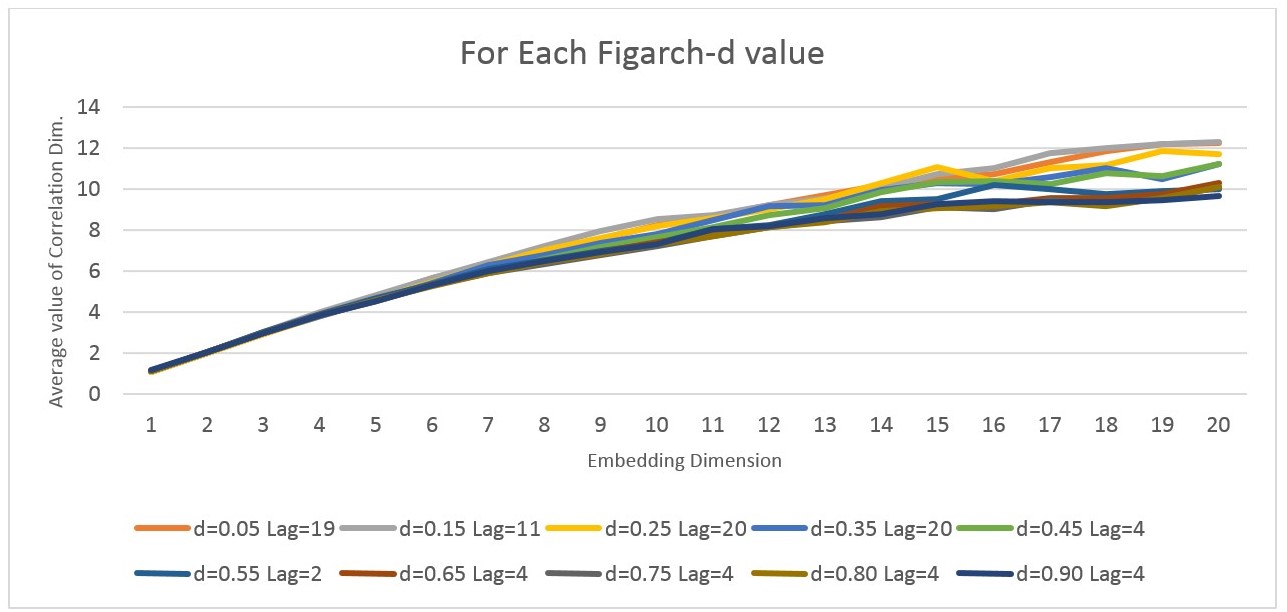}
      \caption{Correlation dimension results according to embedding delay values selected by looking at the most converged lines for each FIGARCH model.}
\end{figure}
\indent
In order to find out embedding dimensions for Figarch simulations, for the application of correlation dimension and the false nearest neighbor method, Kostelich and Swinney suggest that both methods work well when applied to low dimensional (3 or less) chaotic attractors. However, the convergence of the nearest neighbor method seems better for high dimensional attractors. So, considering high dimensional attractors, FNN results are assumed to be sufficient for further analysis.
\section{Lyapunov Exponent}
The Lyapunov exponent is a parameter characterizing the behavior of a dynamical system. It gives the average rate of exponential divergence from nearby initial conditions. If the Lyapunov exponent is positive, then it is suggested that the system is chaotic; if it is negative, the system will converge to a periodic state; and if it is zero, there is a bifurcation.\\
\indent
While there are several Lyapunov exponents, the largest Lyapunov exponent is the most widely used to test chaotic behavior. When the attractor is chaotic, the trajectories diverge, on average, at an exponential rate characterized by the largest Lyapunov exponent.
\subsection{Wolf's Algorithm}
In 1985, Wolf et al. presented an algorithm which allowed the estimation of non-negative Lyapunov exponents from time series data. Equation below provides computation of maximal Lyapunov exponent in a direct way. 
\begin{equation}
\Lambda_{max}={1\over M t_{evolv}} \sum\limits_{i=0}^M ln {L_{evolv}^{(i)}\over L_0^{(i)}}
\end{equation}
where $L_0$ is the Euclidian distance between nearest neighbors of initial point,  $t_{evolv}$ is fixed evolution time with the same order of magnitude as the embedding delay $L_{evolv}$ is final distance between the evolved points and M is the total number of replacement steps.\\
\indent
After gathering two essential input for accurate computations of Lyapunov exponent, correct embedding delay and embedding dimension, in final step, maximal Lyapunov Exponent is calculated first by using Wolf's algorithm.\\
\indent
As shown in Figure 6, maximal Lyapunov exponents converge to positive values in all Figarch $d$ values suggesting extreme sensitivity to changes in initial conditions which is the indication of chaotic behavior.\\
\begin{figure}[!ht]
  \centering
    \includegraphics[width=0.85\linewidth]{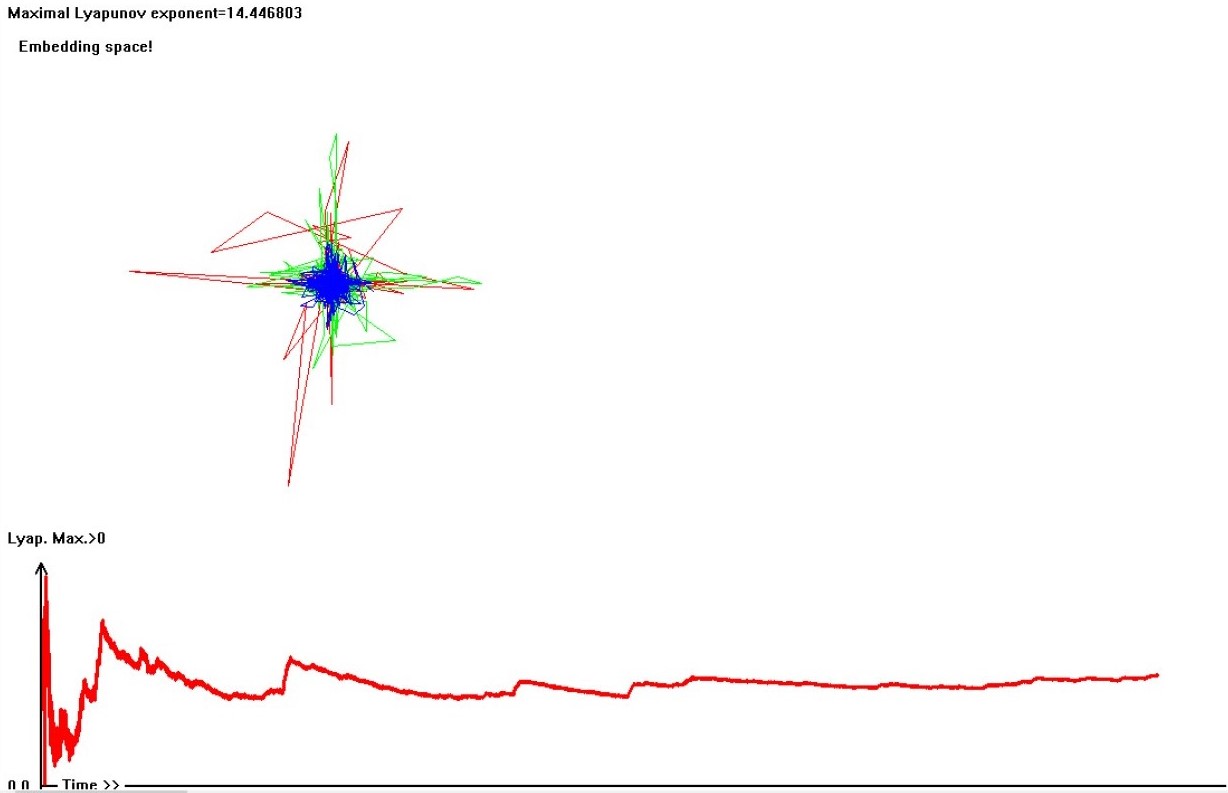}
      \caption{Maximal Lyapunov Exponent result for FIGARCH d=0.90.}
\end{figure}
\begin{table}[h!]
\centering
\begin{tabular}{|c|c|c|c|} 
 \hline
 Figarch Model & Embedding Delay & Embedding Dimension & Max. LE Wolf's Algorithm \\ [0.5ex] 
 \hline
 Figarch d=0.05 & 2 & 6 & 9.1\\ 
 Figarch d=0.15 & 1 & 7 & 2.27\\
 Figarch d=0.25 & 2 & 6 & 11.85\\
 Figarch d=0.35 & 4 & 8	& 2.38\\
 Figarch d=0.45 & 6 & 6 & 7.6\\ 
 Figarch d=0.55 & 7 & 6 & 24.5\\ 
 Figarch d=0.65 & 8 & 6 & 22.17\\
 Figarch d=0.75 & 8 & 6 & 20.8\\ 
 Figarch d=0.80 & 6 & 6 & 21.1\\
 Figarch d=0.90 & 8 & 6 & 14.4\\ [0.5ex]
 \hline
\end{tabular}
\caption{Maximal Lyapunov Exponent results for all FIGARCH d values computed with Wolf's Algorithm.}
\end{table}
\indent
Then, Kantz's algorithm is used to calculate maximal Lyapunov Exponent in order to compare results with Wolf's.
\subsection{Kantz's Algorithm}
Kantz (1994) and Rosenstein (1993) independently proposed a consistent estimator for the maximal Lyapunov exponent. In order to calculate the largest Lyapunov exponent, all neighbors closer to a reference point than given size $\varepsilon$ is identified and average distance of all trajectories to the reference trajectory is calculated.
\begin{equation}
S(t)=\frac{1}{N}\sum\limits_{i=1}^N ln\bigg[\frac{1}{|\Omega_i|} \sum\limits_{j\in\Omega_i}^N |p_i-p_j|\bigg]
\end{equation}
\indent
where $p_i$ are embedding vectors and $|\Omega_i|$ is the number of neighbors in the neighborhood $\Omega_i$ of the reference state $p_i$. If $S(t)$ presents a linear increase, the slope of the fitted line can be taken as an estimate of the maximal exponent.\\
\begin{figure}[!ht]
  \centering
    \includegraphics[width=0.85\linewidth]{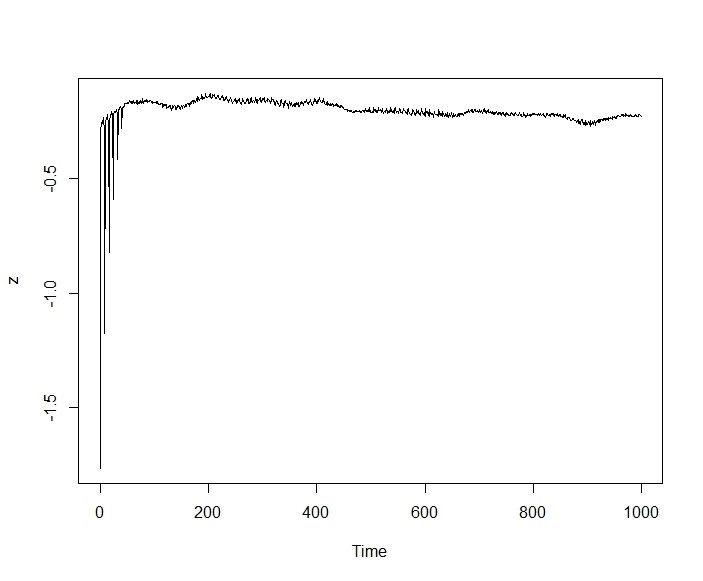}
      \caption{Maximal Lyapunov Exponent result for FIGARCH d=0.90.}
\end{figure}
\begin{table}[h!]
\centering
\begin{tabular}{|c|c|c|c|} 
 \hline
 Figarch Model & Embedding Delay & Embedding Dimension & Max. LE Kantz's Alg. \\ [0.5ex] 
 \hline
 Figarch d=0.05 & 2 & 6 & -3.38E-05\\ 
 Figarch d=0.15 & 1 & 7 & -6.40E-05\\
 Figarch d=0.25 & 2 & 6 & -7.47E-05\\
 Figarch d=0.35 & 4 & 8	& -7.93E-05\\
 Figarch d=0.45 & 6 & 6 & -8.19E-05\\ 
 Figarch d=0.55 & 7 & 6 & -3.50E-06\\ 
 Figarch d=0.65 & 8 & 6 & -0.00011405\\
 Figarch d=0.75 & 8 & 6 & -0.000119346\\ 
 Figarch d=0.80 & 6 & 6 & -6.19E-05\\
 Figarch d=0.90 & 8 & 6 & -0.000124053\\ [0.5ex]
 \hline
\end{tabular}
\caption{Maximal Lyapunov Exponent results for all FIGARCH d values computed with Kantz's Algorithm.}
\end{table}
\indent
Contrary to Wolf's algorithm results, calculations with Kantz's algorithm produce negative values for  maximal Lyapunov exponents for all figarch d values contradicting that FIGARCH is a deterministic chaotic system.
\subsection{Direct Approach by Constructing Dimensional Map}
Finally, we compute maximal Lyapunov Exponent by using the dynamical rules of the map directly from the equation rather than simulated data for which we made calculations with Wolf's and Kantz's algorithm.\\
\indent
Phase space trajectory is computed for Figarch $\{x_1 (i),y_1 (i)\}$, for i=1 to 5000. Local Lyapunov exponent at every point is computed along the trajectory and then all local Lyapunov exponents are averaged. Each point on the trajectory and its neighbor can be written as;
\begin{equation}
\{x_1(i),y_1(i)\}\to\{x_1(i+1),y_1(i+1)\}
\end{equation}
\begin{equation}
\{x_2(i),y_2(i)\}\to\{x_{2t}(i+1),y_{2t}(i+1)\}
\end{equation}
\indent
Local Lyapunov exponent is;
\begin{equation}
LLE=\frac{\sqrt{(x_{2t}(i+1)-x_1(i+1))^2+(y_{2t}(i+1)-y_1(i+1))^2}}{\sqrt{(x_2(i)-x_1(i))^2+(y_2(i)-y_1(i))^2}}
\end{equation}
\indent
Besides, using $\{x_{2t} (i+1),y_{2t} (i+1)\}$ as starting point for next iteration will cause $\{x_2 (i),y_2 (i)\}$ trajectory part more in each iteration. So, contracting $\{x_{2t} (i+1)-x_1 (i+1),y_{2t} (i+1)-y_1 (i+1)\}$ by $(d(i))\over(dt(i))$ will keep the starting point of the next iteration is as close to the trajectory as it was on the previous one.\\
\begin{figure}[!ht]
  \centering
    \includegraphics[width=0.85\linewidth]{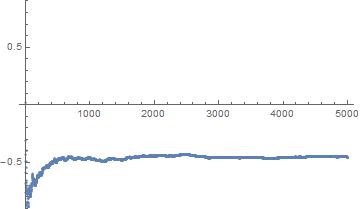}
      \caption{Maximal Lyapunov Exponent results for FIGARCH d=0.80.}
\end{figure}
\begin{table}[h!]
\centering
\begin{tabular}{|c|c|c|c|} 
 \hline
 Figarch Model & Max. LE Wolf's Alg. & Max. LE Kantz's Alg. & 2 Dimensional Map\\ [0.5ex] 
 \hline
 Figarch d=0.05 & 9.1 & -3.38E-05 & -1.05824 \\ 
 Figarch d=0.15 & 2.27 & -6.40E-05 & -0.375591\\
 Figarch d=0.25 & 11.85 & -7.47E-05 & -0.195386\\
 Figarch d=0.35 & 2.38 & -7.93E-05 & -0.159978\\
 Figarch d=0.45 & 7.6 & -8.19E-05 & -0.18594\\ 
 Figarch d=0.55 & 24.5 & -3.50E-06 & -0.249452\\ 
 Figarch d=0.65 & 22.17 & -0.00011405 & -0.326646\\
 Figarch d=0.75 & 20.8 & -0.000119346 & -0.409031\\ 
 Figarch d=0.80 & 21.1 & -6.19E-05 & -0.455532727\\
 Figarch d=0.90 & 14.4 & -0.000124053 & -0.616878\\ [0.5ex]
 \hline
\end{tabular}
\caption{Maximal Lyapunov Exponent results for Wolf's Algortihm, Kantz's Algorithm and 2 Dimensional Map.}
\end{table}
\indent
To sum up, results here matches with the findings with Kantz's algorithm by delivering negative Lyapunov Exponent values and suggest that FIGARCH models are not deterministic chaotic.
\section{Conclusion}
The correlation dimension and Lyapunov exponents have been calculated to investigate the chaoticity properties of FIGARCH stochastic difference equations. Numerical simulations of the latter served to computations of embedding dimension and delay which in turn are used for the correlation dimension and Lyapunov exponents. Correlation dimension is fractional for all cases which pinpoints the self-similar (fractal) properties of the FIGARCH. However, Wolf's algorithm led to positive Lyapunov exponents for all cases.\\
\indent
Based on these results, we could have concluded that chaotic behavior prevails. This might have been the artifact of the embedding method as it has been expounded in (Dechert and Gencay, 2000). This is to say that the largest Lyapunov exponent may not be preserved under Takens' embedding theorem.\\
\indent
In order to check the Wolf's algorithm, we have also employed Kantz's method and Jacobian derivative based on the stochastic difference equations. Both methods led to negative Lyapunov exponents for all possible cases. The latter indicates the presence of some kind of nonlinear determinism in the data which stemmed from FIGARCH but prohibits the possibility of deterministic chaos. When the data is modelled by FIGARCH stochastic difference equations it is not right to attribute the irregurality and self-similarity to deterministic chaos.
\section{References}
Abarbanel, H. D. 1996 Analysis of Observed Chaotic data. New York: Springer.\\[0.2in]
Baillie, R. T., Bollerslev, T., Mikkelsen, H. O. 1996. Fractionally integrated generalized autoregressive conditional heteroskedasticity. Journal of Econometrics, 74. \\[0.2in]
Baillie R. T., Han Y. W., Myers R. J. 2007 Long Memory and FIGARCH Models for Daily and High Frequency Commodity Prices. Working paper No.594.\\[0.2in]
Brooks, C. 1998. Chaos in foreign exchange markets: a sceptical view. Computational Economics, 11 (3). pp. 265-281.\\[0.2in]
Caporin, M. 2002. FIGARCH models: Stationarity, Estimation Methods and the Identification Problem, Working Paper.\\[0.2in]
Cujaeiro, D. O., Tabak, B. M. 2008. Testing for long-range dependence in world stock markets. Chaos, Solitons and Fractals 37, 918-927.\\[0.2in]
Das, A., Das, P. 2006. Does composite index of NYSE represents chaos in the long time scale?. Applied Mathematics and Computation 174, 483–489.\\[0.2in]
Das, A., Das, P. 2007. Chaotic analysis of the foreign exchange rates Applied Mathematics and Computation. 185, 388–396.\\[0.2in]
Dechert W.D., Gençay, R. 2000. Is the largest Lyapunov exponent preserved in embedded dynamics? Physics Letters A 276, 59–64.\\[0.2in]
Fraser, A. M. and Swinney, H. L. 1986 Independent coordinates for strange attractors from mutual information Phys. Rev. A 33 1134-40.\\[0.2in]
Francq, C. and Zakoian, J.M. 2010. GARCH Models Structure, Statistical Inference and Financial Applications. Wiley.\\[0.2in]
Frezza, M. 2014. Goodness of fit assessment for a fractal model of stock markets. Chaos, Solitons and Fractals 66, 41-50.\\[0.2in] 
Gunay, S. 2015. Chaotic Structure of the BRIC Countries and Turkey’s Stock Market.\\[0.2in]
International Journal of Economics and Financial Issues, 5(2), 515-522.\\[0.2in]
Kantz, H. 1994. A robust method to estimate the maximal Lyapunov exponent of a time series Phys. Lett. A 185 77–87.\\[0.2in]
Kantz, H. and Schreiber, T. 2003. Nonlinear Time Series Analysis. Cambridge University Press.\\[0.2in]
Kaplan, D. T. and Glass, L. 1992. Direct test for determinism in a time series. Phys. Rev. Lett. 68 427-30.\\[0.2in]
Kennel, M.B., Brown, R., Abarbanel, H.D.I. 1992. Determining embedding dimension for phase-space reconstruction using a geometrical reconstruction. Physical Review A, V.45, No.6.\\[0.2in]
Kodba, S., Perc, M., Marhl, M. 2005. Detecting Chaos from a Time Series. European Journal of Physics 26 (2005) 205-215.\\[0.2in]
Kostelich E.J., Swinney H. L. 1989. Practical Considerations in Estimating Dimension from Time Series Data, Physica Scripta. Vol.40, 436-441.\\[0.2in]
Liu, H.F., Dai, Z.H., Li, W.F., Gong, X., Yu, Z.H. 2005. Noise robust estimates of the largest Lyapunov exponent, Physics Letters A 341, 119-127.\\[0.2in]
Moeini A., Ahrari M., Madarshahi S.S. 2007. Investigating Chaos in Tehran Stock Exchange Index. Iranian economic review, 18, 103-120.\\[0.2in]
Rosenstein, M. T., Collins, J.J., and Carlo, J.D.L. 1992. A Practical Method for Calculating Largest Lyapunov Exponents from small data sets.\\[0.2in]
Sviridova, N., Sakai, K. 2015. Human photoplethysmogram: new insight into chaotic characteristics. Chaos, Solitons and Fractals 77, 53-63.\\[0.2in] 
Takens, F. 1981. Detecting strange attractors in turbulence. In Dynamical Systems and Turbulence, Rand D, Young L (eds). Springer-Verlag: Berlin.\\[0.2in]
Wolf, A, Swift, J.B., Swinney, H, Vastano, J. 1985. Determining Lyapunov exponents from a time series. Physica D 16: 285–317.\\[0.2in]
Software packages:\\[0.2in]
https:www.kevinsheppard.com/MFE\_Toolbox\\[0.2in]
http://www.oxmetrics.net/index.html\\[0.2in]
http://www.matjaperc.com/ejp/time.html\\[0.2in]
R packages (fractal, tseriesChaos, RTisean)\\[0.2in]

\end{document}